\documentclass[twocolumn,showpacs,preprintnumbers,amsmath,amssymb,citeautoscript,prb]{revtex4-1}

\usepackage{graphicx}
\usepackage{dcolumn}
\usepackage{bm}
\usepackage{times}

\begin{document}

\title{Contrasts in electron correlations and inelastic scattering between LiFeAs and LiFeP revealed by charge transport}

\author{S.~Kasahara$^{1,2}$}
\author{K.~Hashimoto$^{1}$}
\author{H.~Ikeda$^{1}$}
\author{T.~Terashima$^{2}$}
\author{Y.~Matsuda$^{1}$}
\author{T.~Shibauchi$^{1}$}

\affiliation{
$^{1}$Department of Physics, Kyoto University, Kyoto 606-8502, Japan
\\
$^{2}$Research Center for Low Temperature and Materials Sciences, Kyoto University, Kyoto 606-8501, Japan
}

\date{\today}

\begin{abstract}
By using high-quality single crystals, we quantitatively compare the transport properties between LiFeAs and LiFeP superconductors with compensated electron and hole carriers. The low-temperature resistivity follows the Fermi-liquid $AT^2$ dependence with a factor of $\sim 3$ difference in the coefficient $A$. This highlights weaker electron correlations in LiFeP, which is consistent with its $\sim 70$ times lower upper critical field than that of LiFeAs. Our analysis of the magneto-transport data indicates that in LiFeP the electron carriers with lighter masses exhibit stronger temperature dependence of inelastic scattering rate than the holes, which is the opposite to the LiFeAs case. This stark difference in the band-dependent inelastic scattering may be relevant to the recently reported contrasting superconducting gap structures in these two superconductors.
\end{abstract}

\pacs{
74.70.Xa 
74.25.F- 
74.25.Jb 
74.62.Bf 
}  
\maketitle


A central issue in the physics of iron-based superconductivity \cite{Stewart11,Hirschfeld11} concerns the origin of the pairing interaction, in which the importance of the interband fluctuations associated with spin/orbital degrees of freedoms has been discussed \cite{Mazin08,Kuroki09,Chubukov09,Graser09,Ikeda10,Thomale11,Kontani10}. Among various peculiar properties caused by the multiband electronic structure with good nesting between the hole and electron Fermi surface sheets, the non-universality of the superconducting gap structure is one of the outstanding features in this new class of materials \cite{Hirschfeld11,Hashimoto10,Hashimoto11}. Understanding what causes the nodal and nodeless superconducting gap is believed to be a key to the mechanism of this intriguing superconductivity. It has been theoretically pointed out that the frustration between the electron-hole interband interaction and the electron-electron (intraband) scattering can induce the nodal state \cite{Kuroki09}. It is therefore of primary importance to quantitatively determine the differences in the band-dependent inelastic scattering and in the strength of electron correlations between nodal and nodeless iron-based superconductors.

The normal-state transport properties provide the most fundamental information on the scattering mechanism of charge carriers.  In particular, the masses and scattering rates of the carriers are seriously modified when electron correlations are significant.   Among iron-pnictide families, the transport properties have been studied extensively in the 122 family, where high-quality crystals can be obtained \cite{Stewart11}.  One of the most studied is the $A$Fe$_{2}$(As$_{1-x}$P$_{x}$)$_2$ ($A=$\,Ba, Sr, Ca) system \cite{Kasahara10, Kasahara11}, where the isovalent substitution is found to produce nodal superconductivity \cite{Hashimoto10,Nakai10a,Yamashita11} with relatively high $T_c$ (up to 31\,K) without introducing strong disorder \cite{Shishido09,vanderBeek10}. 
An advantage of this system in the transport studies is that the isoelectronic property of P and As preserves the compensation condition where the number of electron carriers is the same as that of holes, which enables us to analyze the transport data in a much simpler form than the electron- or hole-doped pnictides. In this system, however, the nodal state is found to be robust against P substitution \cite{Hashimoto11,Qiu11}, and thus one cannot directly obtain information about the relationship between the superconducting gap structure and carrier scattering mechanism in the normal state.

Among various iron-pnictides, the 111-family compound LiFeAs ($T_c \approx 18$\,K) \cite{Wang08,Tapp08,Pitcher08} and its counter part LiFeP ($T_c \approx 5$\,K) \cite{Deng09,Mydeen10} provide a unique route to the comparative transport study with keeping the compensation. 
Unlike other stoichiometric iron-arsenides, LiFeAs exhibits neither structural nor magnetic transitions \cite{Wang08,Tapp08,Pitcher08,Song10,Heyer10}, while antiferromagnetic fluctuations have been observed
\cite{Jeglic11,Taylor11}. 
Importantly, recent penetration depth measurements \cite{Hashimoto11} reveal that the low-energy quasiparticle excitations are quite different between LiFeAs and LiFeP; LiFeP has nodal gap structure \cite{Hashimoto11} in contrast to the fully gapped superconducting state in LiFeAs \cite{Kim11,Tanatar11}. 
These features make the detailed comparisons of LiFe$Pn$ ($Pn$\,=\,As or P) as an ideal test to identify the essential connection between the signatures in the inelastic scattering and the structure of superconducting order parameters.

Here, using high-quality single crystals, we make detailed comparisons of the charge transport properties between LiFeAs and LiFeP superconductors. We find that although both compounds exhibit Fermi-liquid transport properties at low temperatures, the strength of the electron correlations is very different. Remarkably, the weaker correlations in nodal LiFeP are in contrast to the general tendency that stronger correlations favor unconventional superconductivity \cite{Hashimoto10b}, in which the strong Coulomb repulsion often gives rise to the gap node formation. The magneto-transport data reveal that these two superconductors have very different temperature dependence of the scattering rates of electrons and holes.  Based on the characteristic features in inelastic scattering of carriers inferred from the transport results, we discuss the origin of the different gap structures in LiFe$Pn$.

\begin{figure}[t]
\includegraphics[width=1.0\linewidth]{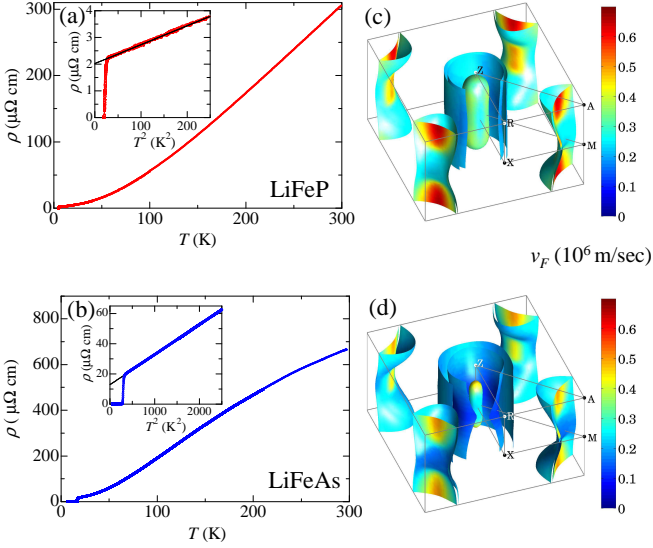}
\caption{(color online). 
In-plane resistivity $\rho_{xx}(T)$ in single crystals of (a) LiFeP and (b) LiFeAs at zero magnetic field. 
The insets show the expanded view plotted against $T^2$ at low temperatures. 
Also shown are the Fermi surface of (c) LiFeP and (d) LiFeAs calculated by the density functional theory \cite{WIEN2k} with the color shades illustrating the magnitude of the Fermi velocity $v_F$. 
} \label{rho-T}
\end{figure}


Single crystals of LiFe$Pn$ are grown by a self flux method using Li (ingots) and FeAs/FeP (powders). Properly weighed starting materials are placed in a BN crucible, and are sealed in a quartz tube. 
The whole materials are heated to 1100$^\circ$C, then slowly cooled down typically to 600$^\circ$C. Single crystals with typical size of 3--4\,mm are obtained for LiFeAs, while the size of LiFeP at this state is up to 150\,$\mu$m.  The gold wires for the transport measurements were attached to the crystals by silver epoxy in the glove box to avoid air exposure. Large residual resistivity ratio ($RRR$) values (see Fig.\:\ref{rho-T} and Table\:\ref{TableI}), along with the recent observations of quantum oscillations \cite{Putzke11}, indicate very small impurity scattering rate in our crystals. Band structure including spin-orbit coupling is calculated by density functional theory implemented in the \textsc{Wien2k} code \cite{WIEN2k} with the experimental lattice parameters in Table\:\ref{TableI}.


Figures\:\ref{rho-T}(a) and (b) show the zero-field in-plane resistivity $\rho_{xx}(T)$ in LiFeP and LiFeAs, respectively. Sharp superconducting transition is seen at $T_{c}$. 
The low-temperature resistivity in the normal state follows $\rho_{xx}(T) = \rho_0 + AT^2$  for both compounds (see the expanded $T^2$-plots in the insets), where $A$ is the Fermi-liquid coefficient. 
This demonstrates the Fermi-liquid transport properties dominated by the electron-electron scattering in these superconductors. It should be noted that our result of LiFeAs is quantitatively consistent with the previous reports \cite{Heyer10,Tanatar11}.
The observed Fermi-liquid transport property in the 111 systems is in contrast to the non-Fermi-liquid $T$-linear $\rho_{xx}(T)$ observed in BaFe$_2$(As$_{1-x}$P$_x$)$_2$ near the magnetic quantum critical point \cite{Kasahara10,Nakai10}, suggesting that both LiFeAs and LiFeP are fairly far from the magnetic instability.
The magnitude of $A=7.0$\,n$\Omega$cm/K$^2$ in LiFeP is nearly three times smaller than $A=20$\,n$\Omega$cm/K$^2$ in LiFeAs.  The $A$ value is closely related to the electronic specific heat coefficient $\gamma$, which measures the effective mass $m^*$, through the relation $A\propto \gamma^2$ \cite{Kadowaki86}.  According to the band-structure calculations, the band masses of both compounds are close in value; the estimated bare $\gamma_b$ values of LiFeAs and LiFeP are 5.5 and 4.3\,mJ/mol K$^2$, respectively. This yields the ratio of $A$ in LiFeAs and LiAsP $\sim 1.6$, which is nearly half of the observed value. In addition, even at the fastest Fermi velocity parts of the electron Fermi surface sheets shown in Figs.\:\ref{rho-T}(c) and (d), which have the largest contribution to the electron transport, the bare mass difference is no more than $\sim30$\%. These results provide strong evidence that the electron correlation is much weaker in LiFeP than in LiFeAs.

\begin{figure}[t]
\includegraphics[width=1.0\linewidth]{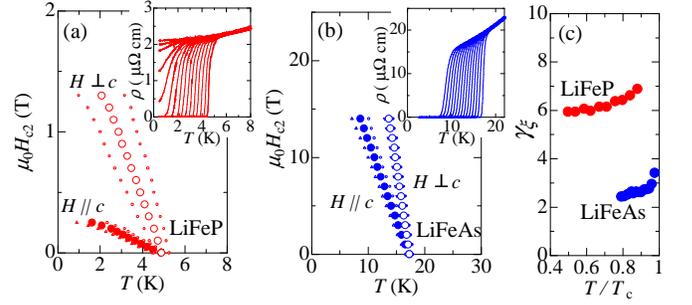}
\caption{(color online). Upper critical field $H_{c2}(T)$ curves in LiFeP (a) and LiFeAs (b) determined from the $\rho_{xx}(T)$ curves under the magnetic field applied parallel to the $c$ axis (insets) and perpendicular to the $c$ axis. For the determination of $H_{c2}(T)$, we used the midpoint of the resistive transition (big circles). $H_{c2}(T)$ determined by using the resistive onset and the zero resistivity point are also shown (small circles). (c) $H_{c2}$ anisotropy $\gamma_{\xi}=H_{c2}(\perp c)/H_{c2}(\parallel c)=\xi_{ab}/\xi_c$ as a function of temperature.
} \label{Hc2}
\end{figure}

The weaker electron correlations in LiFeP is also supported by the upper critical field $H_{c2}$. The resistive transitions to the superconducting state under several magnetic fields are depicted in the insets of Figs.\:\ref{Hc2}(a) and (b). The initial slope of  $H_{c2}(T)$ at $T_c$  for $H\parallel c$, $(dH_{c2}/dT)|_{T_c}$, which is determined by the midpoint of the resistive transition,  is $-1.6$\,T/K for LiFeAs, which is much larger than $-8.5 \times 10^{-2}$ T/K in LiFeP.  This indicates the factor of $\sim 8.3$ difference in the in-plane coherence length $\xi_{ab}$ (see Table\:\ref{TableI}), which is related to the in-plane Fermi velocity $v_F$ (inversely proportional to the in-plane mass $m^*$) and the superconducting gap $\Delta$ by $\xi_{ab}=\hbar v_F/\pi\Delta$. By considering the gap difference by the $T_c$ ratio in the two compounds, the above estimation also provides another piece of evidence for a factor of $\sim2$ heavier $m^*$ in LiFeAs. Here we note that the $H_{c2}$ anisotropy $\gamma_{\xi}$ shown in Fig.\:\ref{Hc2}(c) is temperature dependent, which is a signature characteristic to the multiband superconductors \cite{Lee10,Kurita11,Zhang11,Konczykowski11}. The larger anisotropy in LiFeP follows the nontrivial trend found in the pressure dependence of $\gamma_{\xi}$ in BaFe$_s$(As,P)$_2$ \cite{Goh10}, which may also be associated with the multigap nature.

\begin{figure}[t]
\includegraphics[width=1.0\linewidth]{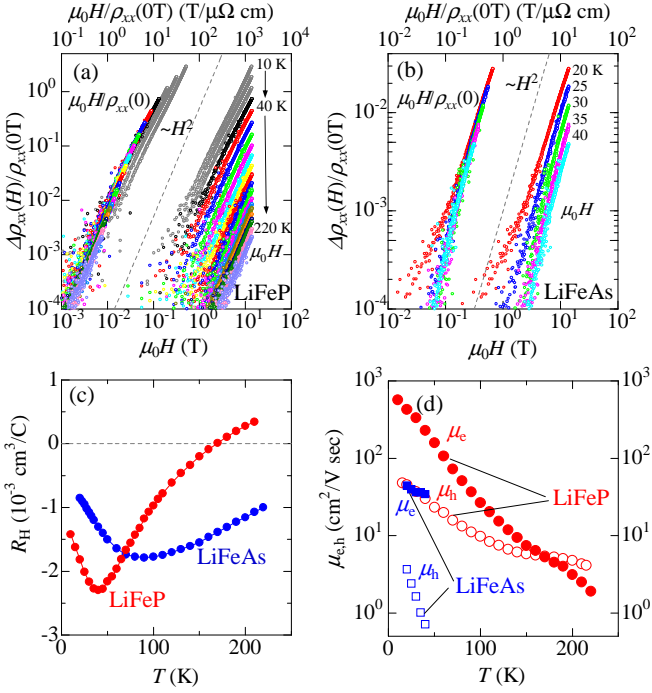}
\caption{ (color online). MR at different temperatures and the Kohler's plot in single crystals of (a) LiFeP and (b) LiFeAs. (c) $R_H(T)$ curves defined by the field derivative of the Hall resistivity $d\rho_{xy}/dH$ in the zero-field limit $H \rightarrow 0$ \cite{note}. 
(d) Electron and hole carrier mobilities $\mu_{e}(T), \mu_{h}(T)$ in LiFeP and LiFeAs derived by using Eqs.\:(\ref{mu_MR}) and (\ref{mu_Hall}). The crossing of the two curves in LiFeP at $T \approx 170$\,K corresponds to the sign change of the Hall effect (c).
} \label{MR}
\end{figure}

The Hall coefficient and magnetoresistance (MR) provide further detailed information about the carrier scattering.  Figures\:\ref{MR}(a) and (b) depict the MR data $\Delta\rho_{xx}(H)/\rho_{xx}(0) \equiv [\rho_{xx}(H) - \rho_{xx}(H=0)]/\rho_{xx}(H=0)$ at several temperatures plotted as  functions of $\mu_0H$ and  $\mu_0H/\rho_{xx}(0)$  for LiFeP and LiFeAs, respectively. In LiFeP, while the MR at the same field  changes almost three orders of magnitude with varying temperature from 40\,K to 220\,K, all the curves collapse onto a single curve when plotted as a function of  $\mu_0H/\rho_{xx}(0)$, indicating that the Kohler's rule is obeyed.  The violation of Kohler's rule is observed below 40\,K.  In LiFeAs the Kohler's rule holds except at very low temperature and low field regime. 
The scaling by $\Delta\rho_{xx}(H)/\rho_{xx}(0) = f(\mu_0H/\rho_{xx}(0)) = f(\omega_c\tau)$ (where $f$ is a function of the cyclotron frequency $\omega_c$ and the scattering time $\tau$) indicates the validity of the simple single-band picture. In the multiband system this suggests that either the contribution from one band (with large $\omega_c\tau$) dominates and other contributions are negligibly small, or all the dominant bands exhibit small and comparable $\omega_c\tau$ values, which mimic the single-band MR. The deviation from the Kohler's rule in LiFeP below $T \alt 40$\,K then indicates the violation of the single-band picture, which appears to require the multiband treatment of the analysis.  This is also supported by the temperature dependence of the Hall coefficient $R_H(T)$  shown in Fig.\:\ref{MR}(c). In LiFeP $R_H(T)$ is positive at $T \agt 170$\,K and becomes negative at lower temperatures.  Below $T \sim 40$\,K where the MR deviates from the single-band Kohler's rule, $R_H(T)$ exhibits a rapid increase. 

Here we analyze the magneto-transport data simply by assuming the two-band model.   In the compensation condition, which is fulfilled in LiFe$Pn$, the low-field magnetoresistance is described as \cite{Pippard}
\begin{equation}
\Delta\rho_{xx}(H)/\rho_{xx}(0) \approx (\omega_c\tau)_e(\omega_c\tau)_h 
= \mu_M^2H^2,
\label{mu_MR}
\end{equation}
where the magnetoresistance mobility $\mu_{M}$ is given by the product of hole and electron mobilities, i.e. $\mu_{M}^2 = \mu_h\mu_e$. 
The Hall mobility $\mu_H$ is given by 
\begin{equation}
\mu_H = R_H/\rho_{xx} = \mu_h - \mu_e.
\label{mu_Hall}
\end{equation}
Thus the combination of the MR and Hall coefficient data in a two-band compensated metal enables us to separate the electron- and hole-mobility $\mu_{h (e)} = e\tau_{h (e)}/m^\ast_{h (e)}$.  The negative Hall coefficient at low temperatures indicates $\mu_e>\mu_h$, which is consistent with larger $v_F$ in the electron sheets from the band-structure calculations in Figs.\:\ref{rho-T}(c) and (d). The extracted temperature dependence of each mobility is shown in Fig.\:\ref{MR}(d). In LiFeP, the temperature dependence of the electron mobility is significantly stronger than that of hole mobility.  On the other hand, in LiFeAs the temperature dependence of the electron mobility is much weaker than that of hole mobility.  This contrasting behavior in the temperature dependence of the carrier mobilities implies a different characteristic feature of the electron inelastic scattering. 

\begin{table*}[t]
 \caption{Comparisons of structural, normal-state, and superconducting properties between LiFeAs and LiFeP based on the present single crystalline study. Lattice parameters of LiFeAs marked with $\dag$ are from Ref.\:\onlinecite{Tapp08}. $h_{Pn}$ is the pnictogen height from the iron plane. Effective masses marked with $\ddag$ (in unit of free electron mass $m_0$) are from the de Haas-van Alphen quantum oscillation measurements \cite{Putzke11}, where the masses for several cyclotron orbits are reported. The mobility ratio $\mu(20\,{\rm K})/\mu(40\,{\rm K})$ represents how strong the temperature dependence of scattering rate $1/\tau(T)$ is for each carriers. The orbital-limiting upper critical field is estimated from the $H_{c2}(T)$ slope (Fig.\:\ref{Hc2}) by the WHH relation $H_{c2}^{\rm WHH}(0) = -0.69T_c\frac{dH_{c2}}{dT}|_{T_c}$ \cite{WHH}. The coherence lengths $\xi_{ab}, \xi_c$ are determined from $H_{c2}^{\rm WHH} (\parallel c)=\Phi_0/2\pi\xi_{ab}^2$ and $H_{c2}^{\rm WHH} (\perp c)=\Phi_0/2\pi\xi_{ab}\xi_c$, where $\Phi_0$ is the flux quantum. For the penetration depth $\lambda_{ab}(0)$ and gap structure marked with $\S$, see Ref.\:\onlinecite{Hashimoto11} and references therein. 
}
 \begin{center}
  \begin{tabular}{lc|cc|c}
    \hline\hline
      & &   LiFeAs & LiFeP  & ratio (LiFeAs/LiFeP)  \\
    \hline
  structural properties & $a$ (\AA) &  3.7914(7)$^\dag$ & 3.6955(7) & 1.026 \\
                        & $c$ (\AA) & 6.364(2)$^\dag$ & 6.0411(19)  & 1.053 \\
                        & $h_{Pn}$ (\AA) &  1.505$^\dag$ & 1.327   & 1.134 \\
                        & Fe-$Pn$-Fe bond angle (deg) &  102.8 & 108.6   & 0.95 \\
    \hline
  normal-state properties & $\rho({\rm 300\,K})$ $(\mu\Omega$cm) & 690 & 310 &  2.2  \\
&$RRR=\rho({\rm 300\,K})/\rho_0$& 53 & 150 & 0.34 \\
                          & $A$ (n$\Omega$cm/K$^2$)            & 20  & 7.0 & 2.9 \\
                          & $m_e^{\ast}$ ($m_{0}$) & 5.4--6.3$^\ddag$ & 2.2--3.6$^\ddag$ & 1.8--2.6 \\
                          & $m_h^{\ast}$ ($m_{0}$) &  -- & 1.0--7.7$^\ddag$ & -- \\
                          & $\mu_e$(20\,K)/$\mu_e$(40\,K)          &  1.3 & 1.9 & 0.69 \\
                          & $\mu_h$(20\,K)/$\mu_h$(40\,K)          &  5.2 & 1.5 & 3.5 \\
    \hline
  superconducting properties & $T_c$ (onset, midpoint) (K) & 17.8, 17.3 & 5.3, 4.9 & 3.4, 3.5  \\
                             & $\mu_0 H_{c2}^{\rm WHH}(0) (H\parallel c, H\perp c)$ (T) & 20, 48 & 0.29, 1.7 & 67, 29 \\
                             & $\xi_{ab}(0), \xi_{c}$(0) (nm) & 4.1, 1.6 & 34, 5.8 & 0.12, 0.28 \\
                             & $\lambda_{ab}(0)$ (nm) & 210$^\S$ & $\sim$ 150$^\S$ & $\sim$ 1.4 \\
                             & $\kappa = \lambda_{ab}/\xi_{ab}$ & 51 & $\sim$ 4.4 & $\sim$ 12 \\
                             & gap structure & nodeless$^\S$ & nodal$^\S$ & -- \\
    \hline\hline
  \end{tabular}
 \label{TableI}
 \end{center}
\end{table*}

Table\:\ref{TableI} shows the quantitative comparisons between LiFeAs and LiFeP, obtained by the present transport studies. Both compounds have very similar Fermi surface topology [Figs.\:\ref{rho-T}(c) and (d)]. 
Three hole sheets present in LiFeP, as confirmed by the quantum oscillation experiments \cite{Putzke11}, particularly suggest that the emergence of nodes in LiFeP is not related with the disappearance of the $d_{xy}$ hole sheet \cite{Kuroki09}. The contrasting temperature dependence of the carrier mobilities between LiFeAs and LiFeP indicates that the carrier scattering processes are significantly different in the two compounds. In the present systems far from the magnetic quantum criticality, the temperature dependence of the carrier mobility $\frac{e\tau}{m^*}(T)$ mainly stems from the temperature dependent inelastic scattering. The fact that in LiFeP the electron mobility has much stronger temperature dependence than the hole mobility implies that the intraband inelastic scattering within electron pockets in LiFeP plays a more important role in the electronic properties than in LiFeAs. The smaller mass enhancement observed for one of the hole sheets in the quantum oscillation experiments is fully consistent with the less significant hole-electron interband scattering in LiFeP \cite{Putzke11}. It has been pointed out that the enhanced electron scattering between the electron pockets leads to the nodal gaps within the electron sheets, which is consistent with the nodal state in LiFeP. 
Clearly, the observed contrasting behaviors of the normal and superconducting properties of LiFeAs/P provide the key to clarifying the pairing mechanism of the iron-pnictides. Further investigations such as determination of the nodal position in LiFeP will help understand the pairing origin of this class of materials.

In summary, we have measured the magneto-transport properties in the high-quality single crystals of LiFeP and LiFeAs. Both compounds show the Fermi-liquid behaviors, which allow us to obtain the strength of electron correlation effect and to separate the electron and hole carrier contributions to the inelastic scattering. The contrasting temperature dependence of the inelastic scattering rate of electrons and holes in the two compounds suggests that the intra-electron scattering within the electron pockets may be important to the formation of nodes in the iron-based superconductors.

We thank A.\,I. Coldea, A. Carrington, H. Kontani, T. Tohyama, Y. Nakai, K. Ishida, Y. Maeno and H. Yamochi for technical help and valuable discussions. 
This work is supported by KAKENHI from JSPS, Grant-in-Aid for GCOE program ``The Next Generation of Physics, Spun from Universality and Emergence'' from MEXT, Japan. 



\begin{thebibliography}{99}
\bibitem{note}
In LiFeP $\rho_{xy}$ is not linear in $H$ at low temperatures, reflecting the large $\omega_c\tau$ values.
\end{thebibliography}
\end{document}